# Generating Primes Using Partitions

Ganesh Reddy Pittu

**Abstract**: This paper presents a new technique of generating large prime numbers using a smaller one by employing Goldbach partitions. Experiments are presented showing how this method produces candidate prime numbers that are subsequently tested using either Miller Rabin or AKS primality tests.

**Introduction**

Generation of large prime numbers is fundamental to modern cryptography protocols [1],[2], generation of pseudorandom sequences [3]-[5], and in new application of these protocols to multi-party computing and cryptocurrencies [6]-[8]. Public-key cryptography requires random generation of prime numbers to derive public key. This paper presents a new technique of generating large prime numbers using a smaller one by employing Goldbach partitions [10]. The algorithm is described and experiments are presented showing how this method gives large primes in an effective manner. A candidate prime will be tested using either Miller-Rabin (MR) or AKS primality tests [11],[12].

**Generation of random primes**

Large prime numbers are generated randomly by considering a random number and testing it with MR or AKS primality test or one might use different sieves [13]-[18], with applications to a variety of cryptography areas (e.g. [19],[20]). The table below presents the average of 10 executions for random generation of prime numbers in a typical experiment.

Table 1. Average attempts to generate a random prime

| Length of the Random Prime number | Average attempts to generate a prime number |
|---|---|
| 45 | 98 |
| 50 | 159 |
| 55 | 172 |
| 60 | 211 |
| 70 | 224 |

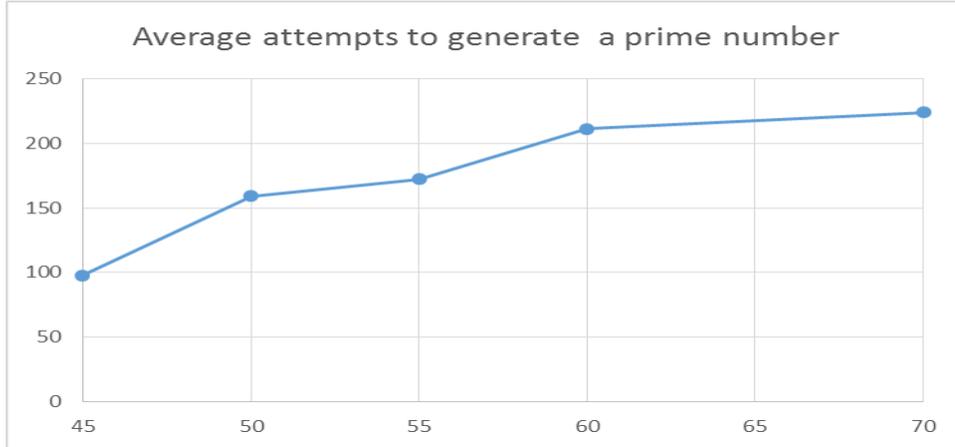

Figure 1. Average of 10 executions to generate a random prime number vs. Length of the prime number. X-axis: Length of the prime number, Y-Axis: Average Attempts.

The average number for 10 executions increases with the size of the number, and it ranges from about 100 for a 45-digit number to 224 for a 70-digit number.

**Generating primes using Goldbach partitions**

The method of prime number generation using Goldbach partitions is now presented. Let $g(n)$ be the number of unique ways $n$ can be expressed as p + q, where p and q are primes. The partition function $g(n)$ has a local maximum for multiples of primes. We obtain local maximum at multiples of primes such as 2x3 = 6; 2x3x5 = 30; 2x3x5x7 = 210; 2x3x5x11 = 330; 2x3x5x13 = 390; 2x3x5x17= 510; 2x3x5x19 = 570; 2x3x5x7x11 = 2310; 2x3x5x7x11x13 = 30030; 2x3x5x7x11x13x17 = 510510, etc. For n=, prime numbers in the range [$n/2$, $n-2$] are among the partitions. The number $n=210$ is the largest number where all numbers in Goldbach partitions are prime numbers.

Example for $n=30$: the Goldbach partitions are (23, 7), (19, 11) and (17, 13).
Example for n=210: the Goldbach partitions are (199,11), (197,13), (193,17), (191,19), (181,29), (179,31), (173, 37), (167, 43), (163, 47), (157, 53), (151, 59), (149, 61), (139, 71), (137, 73), (131, 79), (127, 83), (113, 97), (109, 101), (107, 103). If one counted from the smaller prime up then (187,23) is one pair that is left out since 187 is not prime.

Since we are interested in generating candidate primes, we define $h(n)$ as the count of all primes in the range [$n/2$, $n-2$].

$h(n) \geq g(n)$.

Equality is obtained only for n=6, 30, and 210.

Table 2 presents the number of candidate partitions in the range [$n/2$, $n-2$] obtained by the product of prime numbers for some values of $n$.

Table 2. Number of candidate partitions, $h(n)$

| Prime Numbers | $n$ | No. of Candidate Partitions $h(n)$ | Prime Numbers | $n$ | No. of candidate Partitions $h(n)$ |
|---|---|---|---|---|---|
| 2,3,5 | 30 | 3 | 2,3,5,23 | 690 | 56 |
| 2,3,7 | 42 | 4 | 2,3,7,17 | 714 | 56 |
| 2,3,11 | 66 | 7 | 2,3,7,19 | 798 | 60 |
| 2,3,13 | 78 | 9 | 2,3,11,13 | 858 | 65 |
| 2,3,17 | 102 | 10 | 2,3,5,29 | 870 | 66 |
| 2,3,19 | 114 | 13 | 2,3,5,31 | 930 | 67 |
| 2,3,23 | 138 | 13 | 2,3,7,23 | 966 | 70 |
| 2,3,29 | 174 | 16 | 2,3,5,37 | 1110 | 84 |
| 2,3,31 | 186 | 18 | 2,3,11,17 | 1122 | 85 |
| 2,3,5,7 | 210 | 19 | 2,3,7,29 | 1218 | 87 |
| 2,3,5,11 | 330 | 28 | 2,3,5,41 | 1230 | 87 |
| 2,3,5,13 | 390 | 32 | 2,3,11,19 | 1254 | 90 |
| 2,3,7,11 | 462 | 38 | 2,3,5,7,11 | 2310 | 151 |
| 2,3,5,17 | 510 | 42 | 2,3,5,7,13 | 2730 | 179 |
| 2,3,7,13 | 546 | 42 | 2,3,5,7,17 | 3570 | 223 |
| 2,3,5,19 | 570 | 42 | 2,3,5,7,19 | 3990 | 248 |
| 2,3,7,11,13 | 6006 | 352 | 2,3,7,11,17 | 7854 | 447 |
| 2,3,7,11,19 | 8778 | 496 | 2,3,11,13,17,19 | 277134 | 11323 |
| 2,3,7,13,17 | 9282 | 522 | 2,3,11,13,17,23 | 335478 | 13517 |
| 2,3,7,13,19 | 10374 | 582 | 2,3,11,17,19,23 | 490314 | 19148 |

The graphs below compares the value of *n* and the number of candiate partitions computed.

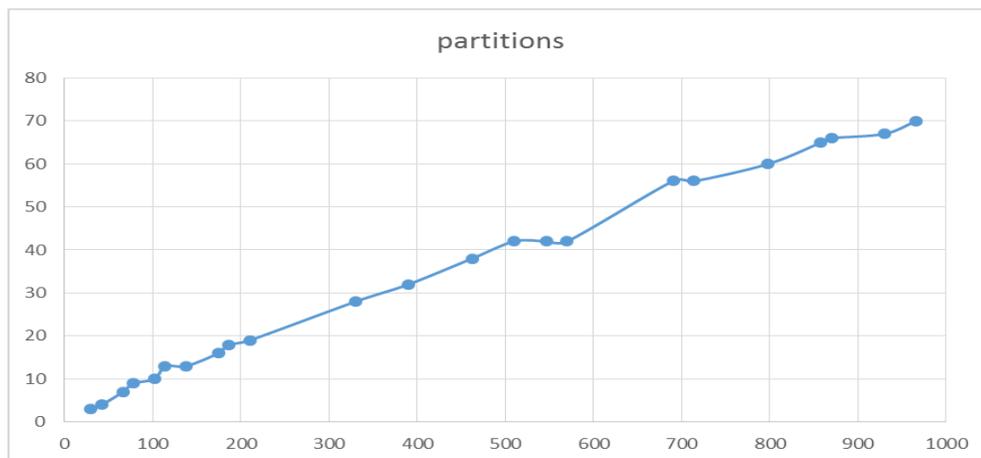

Figure 2. No. of partition, h(n), for *n*<1000

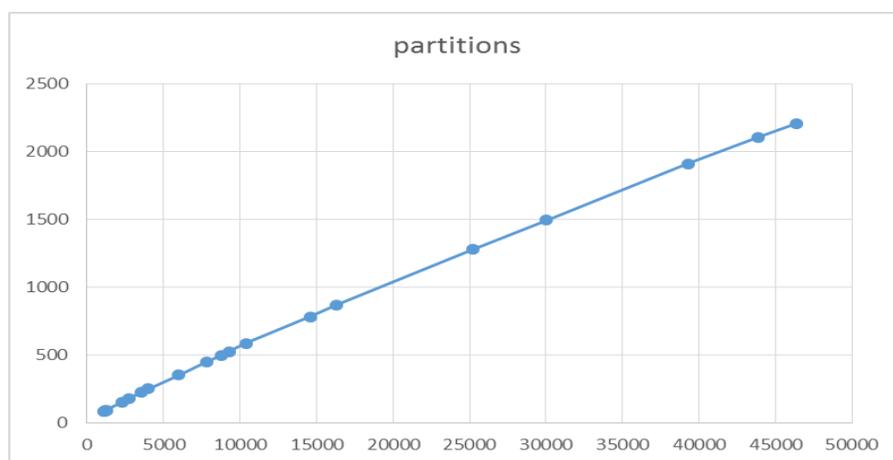

Figure 3  No. of partitions, h(n), for 1000<*n*<50000

Using this technique, we investigate percentage of prime numbers obtained from the partitions by MR and AKS primality testing algorithms.

Table 3 Percentage of Prime Partitions

| Prime Product | Percentage of Prime Partitions | Prime Product | Percentage of Prime Partitions |
|---|---|---|---|
| 30 | 100 | 714 | 66.01 |
| 42 | 80 | 798 | 63.30 |
| 66 | 85.71 | 858 | 65.60 |
| 78 | 77.77 | 870 | 69.69 |
| 102 | 80 | 930 | 64.17 |
| 114 | 76.92 | 966 | 64.28 |
| 138 | 61.53 | 1110 | 64.28 |
| 174 | 68.75 | 1122 | 55.29 |
| 186 | 72.22 | 1218 | 56.32 |
| 210 | 100 | 1230 | 62.50 |
| 330 | 85.71 | 1254 | 56.66 |
| 390 | 84.37 | 2310 | 75.49 |
| 462 | 73.68 | 2730 | 71.50 |
| 510 | 76.19 | 3570 | 69.05 |
| 546 | 71.42 | 3990 | 65.72 |
| 570 | 73.80 | 6006 | 55.39 |
| 690 | 69.64 | 7854 | 52.12 |
| 8778 | 49.19 | 277134 | 33.52 |
| 9282 | 48.65 | 335478 | 32.72 |
| 10374 | 50 | 490314 | 30.64 |

The graphs below presents the percentage of prime numbers generated from the Goldbach partitions.

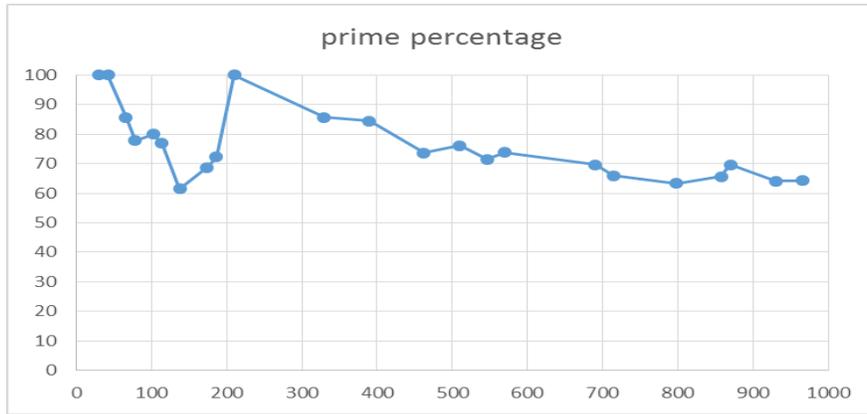

Figure 4. No. of candiate partitions vs percentage of primes for $n<1000$

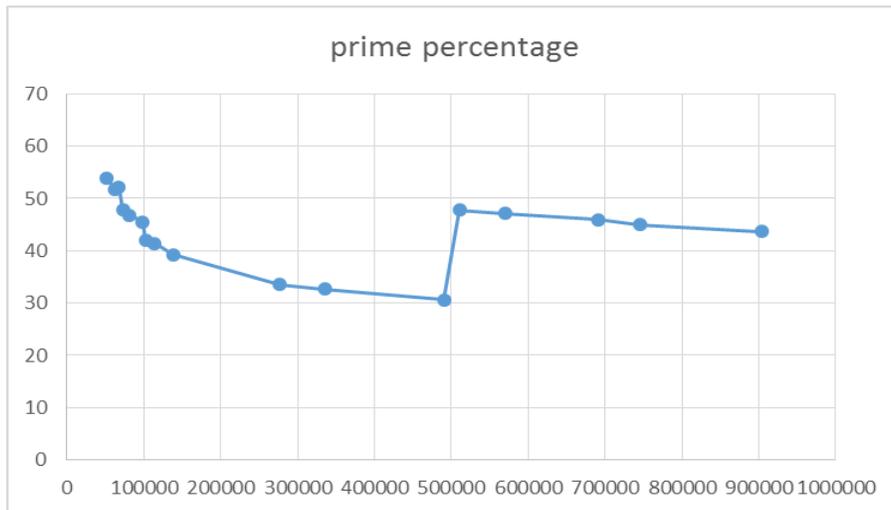

Figure 5. No. of candiate partitions vs percentage of primes for $50000 <n<1000000$

Now we present the new technique of generating a prime number using a smaller one and we call our method the Goldbach Prime Generating Algorithm(GPGA) is presented. This method can be used as an alternate method in public key cryptography and in authentication techniques.

**The GPGA Algorithm:**

1. Randomly pick a large even number $n$.

2. Use a known prime number of certain size, $p_a$, ($p_a << n$) and $p_a + p_b = n$, where $p_b$ is a candidate prime.

3. Test $p_b$ with AKS or MR primality testing algorithms.

4. Repeat 2,3 till $p_b$ is a prime number.

The above algorithm is implemented below.

```
GPGA pseudo code:

n: $\prod_{i=1}^{N} p_i$ , p =2,3,5,7,…..i

Whlie prime == false

        $p_a$ : Known Prime Number ($p_a$ << n)

        $p_b$ : n - $p_a$

        Test $p_b$ for prime using MR or AKS

        If  $p_b$ prime

        Then

                Prime = true
EndWhile

Return $p_b$ .
```

Table 4. GPGA results for the average of 10 Executions

| Length of n (even number) | Length of $p_b$ (known prime) | Average attempts to generate prime number |
|---|---|---|
| 45 | 4 | 21 |
| 45 | 5 | 18 |
| 45 | 6 | 20 |
| 45 | 7 | 15 |
| 45 | 8 | 18 |
| 45 | 9 | 15 |
| 60 | 4 | 10 |
| 60 | 5 | 16 |

| 60 | 6 | 12 |
| 60 | 7 | 18 |
| 60 | 8 | 12 |
| 60 | 9 | 14 |

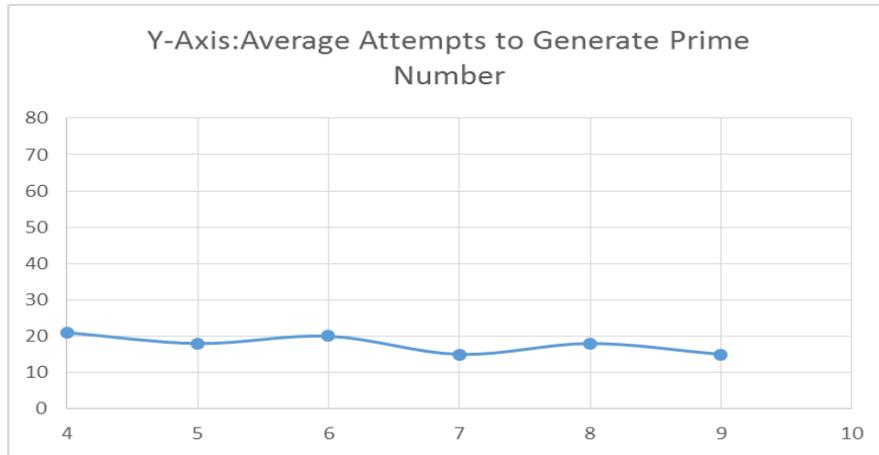

Figure 6. Attempts To Generate Candidate Prime when length of *n = 45*

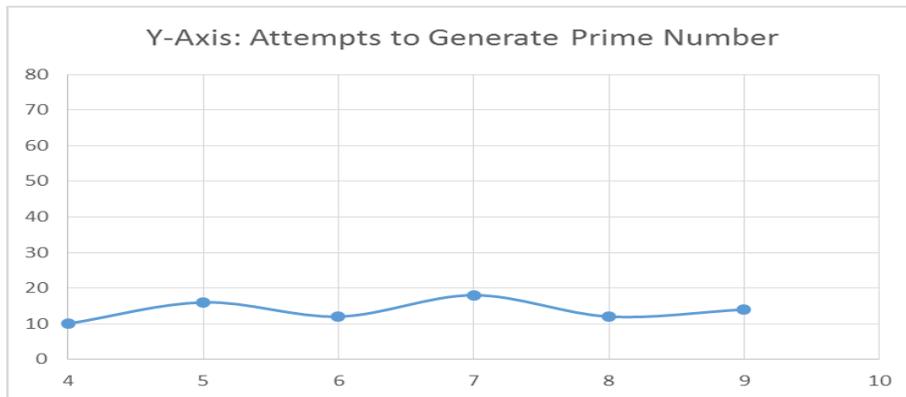

Figure 7. Attempts To Generate Candidate Prime when length of *n = 60*

The above average case is taken for ten executions of the algorithm and are considered for 10 best cases. The while loop iteration might take even more attempts in few cases and will be the worst case. So the best case to obtain a prime number would be O(1).

**Conclusions**
This paper has proposed a new method of generating prime candidates using partitions, which are then tested using primality testing algorithms (MR and AKS). We provide results on the efficiency of this algorithm. The use of the GPGA method reduced the number of trials in the search for primes by a factor of nearly 10 for primes ranging from 45 to 70 digits long.